\documentclass[aps,prl,preprint,amsmath,amssymb,superscriptaddress]{revtex4-1}

\usepackage{graphicx}
\usepackage{epstopdf}
\usepackage{amsmath}

\usepackage{color}

\begin{document}

\title{Single electron gating of topological insulators} 
\author{P.\,Sessi} 
\email{sessi@physik.uni-wuerzburg.de}
\affiliation{Physikalisches Institut, Experimentelle Physik 2, Universit{\"a}t W{\"u}rzburg, 
Am Hubland, 97074 W{\"u}rzburg, Germany}
\author{T.\,Bathon}
\affiliation{Physikalisches Institut, Experimentelle Physik 2, Universit{\"a}t W{\"u}rzburg, 
Am Hubland, 97074 W{\"u}rzburg, Germany}
\author{K.\,A.\,Kokh} 
\affiliation{V.S.\,Sobolev Institute of Geology and Mineralogy, Siberian Branch, 
Russian Academy of Sciences,630090 Novosibirsk,  Russia}
\affiliation{Novosibirsk State University, 630090 Novosibirsk, Russia}
\affiliation{Novosibirsk State University, 630090 Novosibirsk, Russia}
\author{O.\,E.\,Tereshchenko} 
\affiliation{A.V. Rzanov Institute of Semiconductor Physics, Siberian Branch, 
Russian Academy of Sciences, 630090 Novosibirsk,  Russia}
\affiliation{Novosibirsk State University, 630090 Novosibirsk, Russia}
\author{M.\,Bode}
\affiliation{Physikalisches Institut, Experimentelle Physik 2, Universit{\"a}t W{\"u}rzburg, 
Am Hubland, 97074 W{\"u}rzburg, Germany}
\affiliation{Wilhelm Conrad R{\"o}ntgen-Center for Complex Material Systems (RCCM), 
Universit{\"a}t W{\"u}rzburg, Am Hubland, 97074 W{\"u}rzburg, Germany}

\date{\today}

\vspace{1cm}
\begin{abstract}
\vspace{1cm}
Introducing, observing, and manipulating individual impurities coupled to a host material offers the opportunity to create new device concepts based on single spin and charge states. Because of potential applications in spintronics and magneto-electrics, such an approach would be particularly useful for topological insulators (TI), a recently discovered material class hosting spin-momentum-locked surface states. To make them useful for new technologies, a robust control of their interaction with external perturbations is required. However, traditional approaches such as metal electrodes or doping proved to be problematic and resulted in strong mesoscopic fluctuations making the spin-momentum locking ill-defined. Here, we demonstrate the effective gating of TIs by coupling molecules to their surface which, by using electric fields, allow to dynamically control the interface charge state by adding or removing single electrons. This process creates a robust transconductance bistability resembling a single-electron transistor. Our findings make hybrid molecule/TI interfaces functional elements while at the same time pushing miniaturization at its ultimate limit. This opens a new avenue for all electric-controlled spintronic devices based on these fascinating materials.
\vspace{1cm}
\noindent
\end{abstract}


\maketitle
\newpage

\newpage
In the semiconductor industry, the presence of impurities is traditionally considered a limiting factor degrading device performance through the creation of disorder, especially given the ongoing trend towards device miniaturization \cite{RA2005}. Recently, the effects of single dopants were observed in commercial devices \cite{PWJ2010}, severely constraining reproducibility. However, the dramatically improved control of matter at the nanoscale also provided the idea to address single impurities individually \cite{SOK2005,KF2011}. Within this framework, impurities are no longer considered a problem but as an opportunity to push miniaturization to its ultimate level while at the same time creating plenty of room to realize new device concepts, such as single-dopant transistors \cite{FMM2011} or single-spin solid-state quantum computers \cite{ABD2013}.  

In this context, the recently discovered topological insulators (TI) represent a promising new class of materials. TIs are insulating in the bulk but conductive on their surface, where they host linearly dispersing Dirac fermions protected by time-reversal symmetry \cite{KWB2007,HQW2008}. Strong spin-orbit coupling perpendicularly locks the spin to the momentum, resulting in a chiral spin texture which forbids backscattering \cite{RSP2009}. 
All these properties make TIs highly attractive for spintronics, magneto-electrics, and quantum computing \cite{PM2012}. Up to now, however, the controlled gating of topological states, which is required to tune both the spin-texture and the conductance, has proved problematic \cite{Y2013}. In particular, interfacing TIs with external perturbations using traditional approaches such as metal electrodes or dopants  resulted in the undesired creation of strong mesoscopic fluctuations which make the spin-momentum locking ill-defined over length scales of few nanometers \cite{BRS2011,SRB2014}. 

More recently, it has been demonstrated that all these limitations can be overcome at once by using molecules as building blocks \cite{SBK2014,SZY2014,BSK2015,JNS2015,YCL2015,CPL2016}. Molecules have already been successfully used in the past to create hybrid heterostructures with functionalities going beyond those of their constituents \cite{RGB2005,CHW2009}. Especially phthalocyanines (Pc) have been  recently reported to  host several unconventional properties \cite{FSP2011,MRC2012,WHP2015}. In this case, self-assembly processes driven by inherent interactions have been identified as a versatile way to engineer highly homogeneous potential landscapes interfaced with topological states.
\begin{figure}[t] 
	\includegraphics[width=0.6\columnwidth]{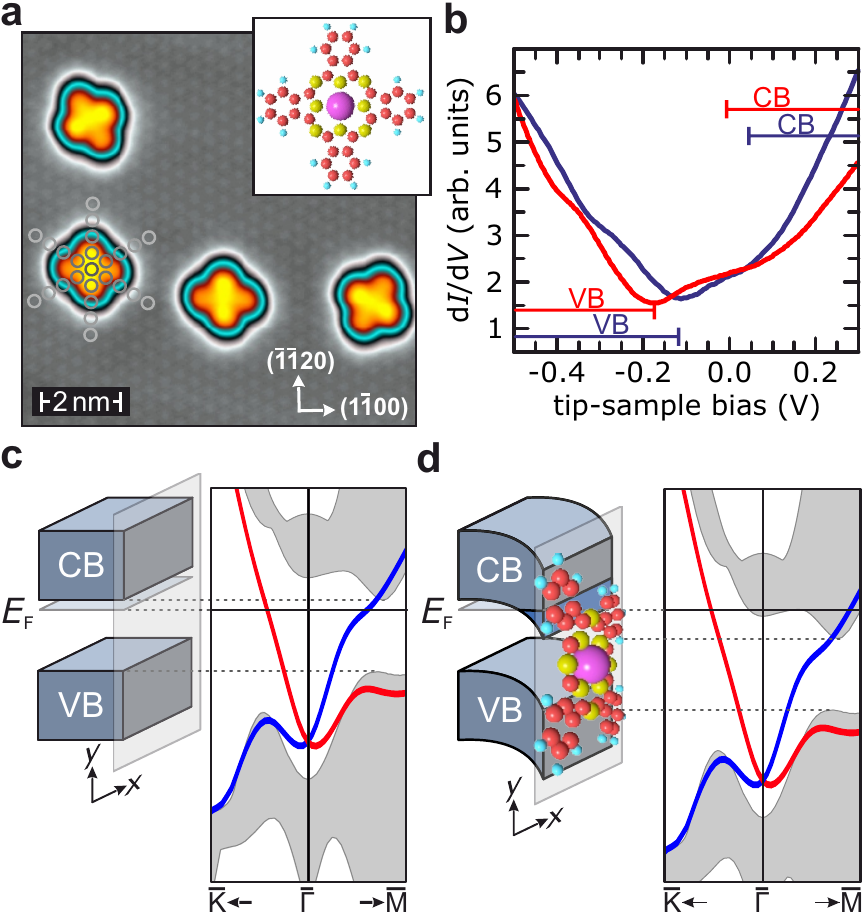}
	\caption{\textbf{a} Constant-current STM image of four MnPc molecules adsorbed on a Bi$_2$Te$_3$(0001) surface. Extrapolation of the surface Te layer reveals an on-top adsorption site. Owing to the substrate's hexagonal surface structure and the 4-fold symmetry of the molecule in the gas phase, three orientations are present onto the surface. Note that, after adsorption,  the molecule symmetry is reduced to 2-fold as a result of the interaction with the  Bi$_2$Te$_3$ substrate. \textbf{b} Charge transfer between the molecule and the substrate leads to an energy shift of the TI that amounts to about 70~meV. Schematic representation of the Bi$_2$Te$_3$ band structure \textbf{c} before and \textbf{d} after molecule deposition causing band bending. Stabilization parameters $V= -0.3$ V, $I = 50$ pA.} \label{Fig1} 
\vspace{0.5cm}
\end{figure}
Here, we demonstrate that the charge state of molecules coupled to TIs can be dynamically switched by using local electric fields. This leads to a transconductance bistability which can be effectively described by a double barrier quantum capacitance (DBQC) effect \cite{L1988}. By spatially mapping the response of the hybrid molecule/TI interface to the charging/discharging process as a function of the applied voltage, a strongly conducting ring of progressively larger diameter develops around each molecule once the ionization threshold is reached, signaling the presence of strong Coulomb repulsions which are only weakly screened by the 2D topological electron gas. Overall, this process results in the gating of the topological insulator surface, which can thus be reproducibly controlled and closely resembles the voltage-current characteristic of a single-electron transistor (SET) device \cite{K1992}. 

Figure 1a shows a dilute concentration of MnPc molecules coupled to the Bi$_2$Te$_3$ surface. All molecules are adsorbed with the same geometry, i.e.\ with the central Mn atom sitting on top of a Te atom, as represented by grey circles interpolating the Te surface lattice to the molecule position. As described in detail in Ref.~14, three different molecular orientations are visible, which result from the combined symmetry of the molecules (4-fold) with that of the surface (6-fold). Comparison of the Bi$_2$Te$_3$ local density of states as inferred by scanning tunneling spectroscopy (STS) before (blue curve in Fig.~1b) and after deposition (red) allows to map the evolution of the electronic properties at the hybrid molecule/TI interface. Subsequent to molecule deposition a moderate negative energy shift is visible, signaling the creation of an interface dipole which bends the Bi$_2$Te$_3$ surface electronic bands, as illustrated in Fig.~1c,d. This observation demonstrates that molecules act as a static potential effectively gating the surface of topological insulators. As described in the following, this has far reaching implications as their charge state can dynamically be controlled by local electric fields. 

\begin{figure}[t] 
	\includegraphics[width=0.70\columnwidth]{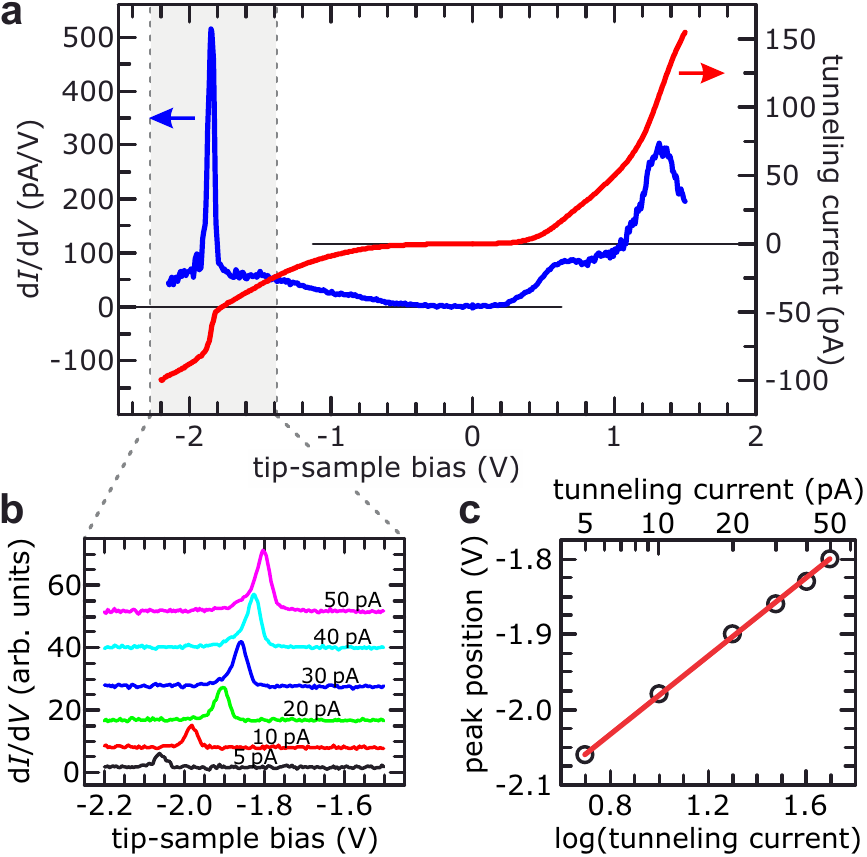}
	\caption{
\textbf{a} $I(V)$ (red line) and $\mathrm{d}I/\mathrm{d}V$ spectrum (blue) measured above the central position of a MnPc molecule. The two broad maxima at positive bias represent empty molecular orbitals. In contrast, the sharp features at about $-1.8$~V are caused by an electric field--driven molecular ionization process. \textbf{b} Current set-point--dependent $\mathrm{d}I/\mathrm{d}V$ spectra measured at the peak position.	Note, that curves are vertically shifted for better visibility. \textbf{c} Plots of the tunneling current dependent position of the ionization threshold and  $d_{z^2}$ orbitals, respectively.  Stabilization parameters $V= -1.5$ V, $I = 30$ pA.
	   \label{Fig2}}  
\vspace{0.5cm}
\end{figure} 
Figure~2a reports the $I(V)$ curve (red line) and the differential conductance $\mathrm{d}I/\mathrm{d}V$ (blue) obtained by positioning the tip over the Pc's central Mn atom. A broad shoulder with onset at $+0.3$~V and a peak centered at $+1.3$~V are visible in the $\mathrm{d}I/\mathrm{d}V$ signal which correspond to a hybrid interface resonance state (IS) and a Mn state of $d_{z^2}$ character, respectively \cite{SBK2014} (see Supplementary Information). In addition, an intensive and very sharp peak shows up at negative bias which does not correspond to any conventional electronic state feature. With the set-point current chosen in Fig.~2a, i.e.\ $I = 50$~pA, it appears at approximately $-1.8$~V. As signaled by a step-like feature in the $I(V)$ curve, its onset is accompanied by a sudden increase of the tunneling current, suggesting that the molecule/TI interface switches to a different state. 

Figure 2b reveals that the position of this sharp and intensive peak strongly depends on the tip--sample distance which can conveniently and precisely be controlled by the STM set-point current. In addition, due to the stronger overlap between the tip's and sample's wave functions the peak height gradually rises. Fig.~2c report the energy positions of the peaks visible in b as a function of the tunneling current flowing through the molecule--TI junction. Since the tunneling current exponentially depends on the barrier width, the linear dependence visible in the semi-logarithmic plot demonstrates that the change in the transconductance is inversely proportional to the the tip-sample separation. 

Within a single impurity picture, these observations can be effectively explained as an electric field--driven reversible ionization process, similar to earlier findings around dopants buried into conventional semiconductors \cite{TVL2008,LG2010} or molecules coupled to insulating substrates \cite{WZC2004,WHP2015}. However, TIs are fundamentally different with respect to these systems since they host on the surface a 2D electron gas which is generally expected to effectively screen electric fields. Our measurements, in line with previous observations on the surface energy level alignment \cite{CSZ2010} and the charge states of bulk dopants \cite{SJW2012}, clearly demonstrates this not being the case. Indeed, as schematically illustrated in Fig.~3a, the electric field partially penetrates into the TI, bending its band structure creating a quantum well at the surface (see Supplementary Information).  Note that the weak screening of topological states as revealed in our experiments has far reaching implications. It proves  that topological states do not behave as a simple metal and imply the existence of an upper limit to their transport properties.

The creation of hybrid molecule-TI interfaces has several advantages, namely: (i) the high degree of charge localization of molecules results in an electronic occupation which can change only by integer numbers, creating a well-defined sequence of discrete states generated by strong Coulomb repulsions, (ii) molecules can be placed onto TI surfaces without perturbing their topological states by scattering \cite{SBK2014}, (iii) contrary to conventional doping, their hybridization with the host is reduced, thus preserving their functionality and (iv) their distribution can be precisely controlled through self-assembly processes \cite{BSK2015}.

Fig.~3b-d displays a more detailed picture how the band diagram of the tip--molecule--TI tunnel junction changes upon application of a negative bias voltage and how this results in an electric field--driven molecular ionization process which eventually acts back on the TI by gating the topological surface state. At zero bias (Fig.~3b) the Fermi levels of tip and sample are aligned. The valence and conduction band edges of the TI are only moderately shifted downwards due to the above-mentioned interface dipole associated to molecule adsorption (cf.\ Fig.~1d). The hybrid interface resonance state (IS) is energetically located at about $+0.3$~eV, i.e.\ in the empty states. As a small negative tip--sample bias voltage is applied (Fig.~3c) the electric field partially penetrates into the TI surface and further bends its valence and conduction band edges down. Electrostatically, this is equivalent to two capacitors connected in series: within this picture one capacitor would represent the tip and the 2D topological electron gas on either side of the vacuum barrier, the other one the quantum well at the TI surface. The voltage partition across the tip--molecule--TI junction is simply defined by the ratio of the respective capacitances (see Supplementary Information). Since the IS is locked to the TI band structure it also shifts downwards, but it still remains above the Fermi level of the TI.

\begin{figure}[t] 
	\includegraphics[width=0.9\columnwidth]{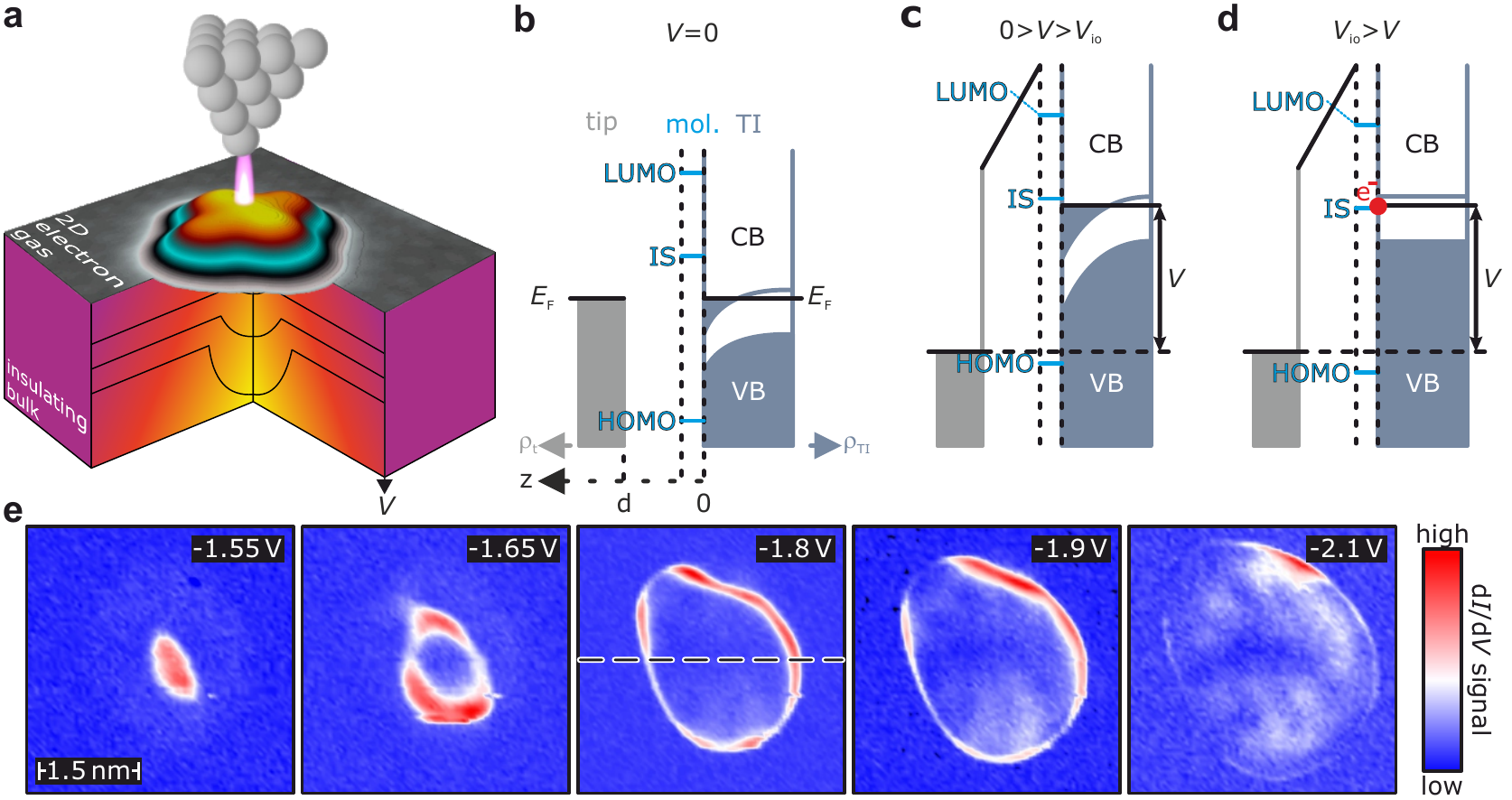}
	\caption{
\textbf{a}~Schematic representation of the tip--molecule--TI tunnel junction with the electric field partially penetrating into the TI. Band diagram of the junction at \textbf{b}~zero bias, \textbf{c}~small negative tip--sample bias, and \textbf{d}~at the threshold voltage leading to a molecular electronic state (IS) that is energetically positioned below the TI Fermi level and therefore charged by an additional electron. \textbf{e}~Series of constant-separation $\mathrm{d}I/\mathrm{d}V$ maps measured at the indicated bias voltages. Stabilization parameters $V= -1.5$ V, $I = 30$ pA.}  \label{Fig3}
\vspace{0.5cm}
\end{figure}
At a certain threshold voltage the band bending is so strong that the IS crosses the Fermi level of the TI and occupation becomes energetically favorable. This situation is sketched in Fig.~3d. The additional electron at the interface creates a repulsive potential which pushes up the surface band structure acting against the band bending (see below). This increases the density of states available for electron tunneling at the surface, 
an effect considered to be particularly important in situations when the tip is away from the molecule and direct tunneling through the conduction channel involving the IS becomes insignificant. Overall, these effects result in the observed step-like increase of the current-voltage relation (cf.\ Fig.~2a) which creates a bistability that can be robustly controlled.

Interestingly, we found that when the tip is displaced laterally from the central Mn atom, its electric field can still effectively gate the surface and ionize the molecule. Fig.~3e shows a bias voltage--dependent series of $\mathrm{d}I/\mathrm{d}V$ maps measured at constant tip--sample separation over the same molecule. The position where the respective threshold bias is reached is signaled by a high $\mathrm{d}I/\mathrm{d}V$ signal. While this threshold bias amounts to about $-1.55$~V at the central position of the molecule,sharp rings with increasing diameters can be observed at more negative bias. Note, that the ring diameter eventually exceeds the size of the molecule. Obviously, the additional, highly localized charge that results from the molecule ionization process cannot be effectively screened by the topological states. As a result the TI is gated, making such an approach a suitable way to manipulate macroscopic transport properties.
\begin{figure}[t!] 
	\includegraphics[width=0.9\columnwidth]{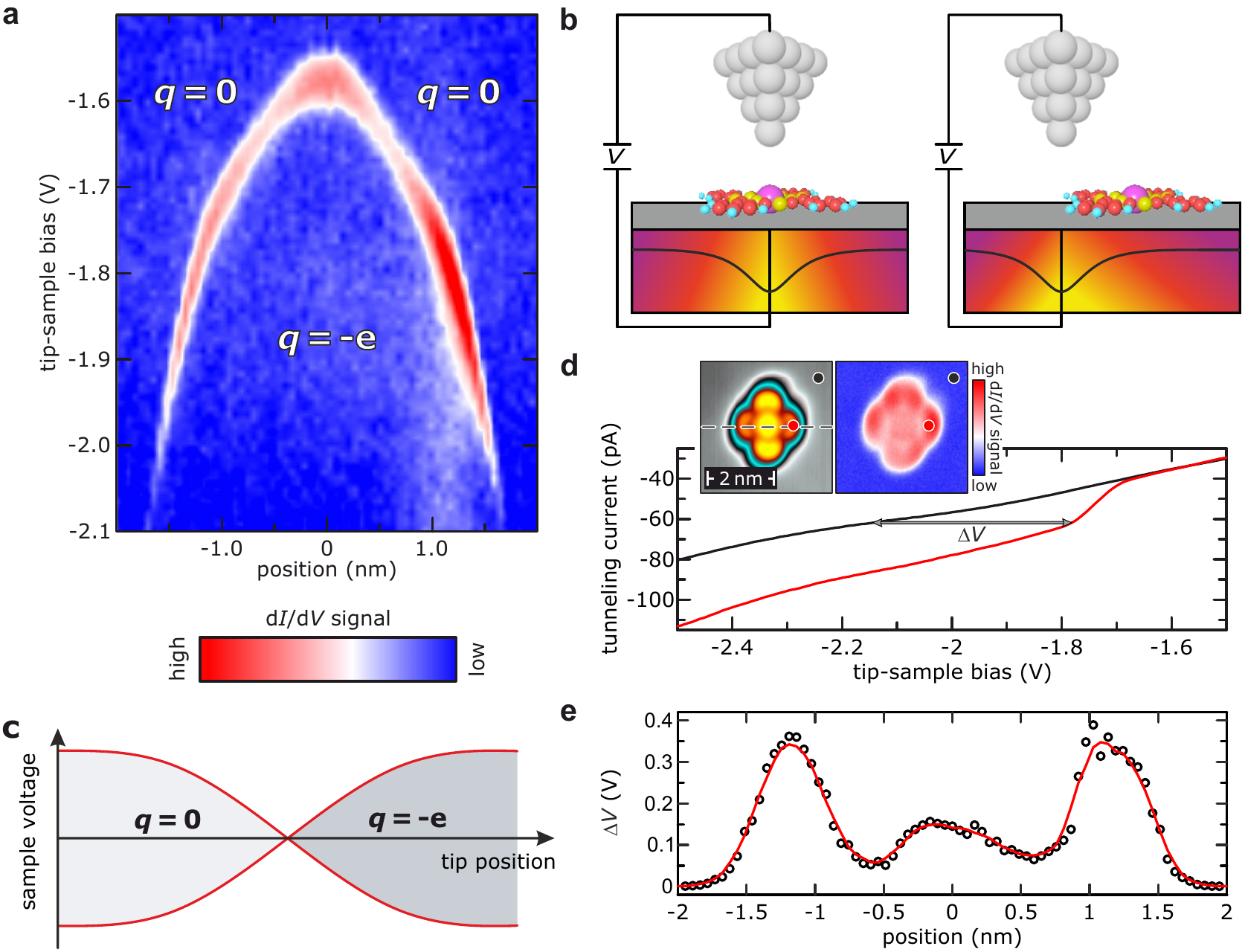}
	\vspace{-0.4cm}
	\caption{
\textbf{a}~Sequence of position-dependent spectra (setpoint parameters: $V = -1.5$~V, $I = 30$~pA) measured along the line drawn in the central panel of Fig.~3e. \textbf{b} Schematics of the tip position dependent band bending: the voltage required for transferring one additional electron at the interface depends on the strength of local electric field underneath the molecule, which is maximized once the tip is positioned exactly on top of it. \textbf{c} Transconductance bistability resulting from the charging discharging process, which depends on the tunneling bias and the molecule gating voltage (in our case adjustable by changing the tip position), thus resembling the current-voltage relation of a SET. \textbf{d}~$I(V)$ spectra taken on the bare TI surface (black) and on the MnPc molecule (red). Precise measurement positions are marked by points in the insets showing the topography and a $\mathrm{d}I/\mathrm{d}V$ map at the energy of the IS ($V = 0.3$~V). The charging of the molecule leads to a sharp increase of the tunneling current. \textbf{e}~Plot of the voltage shift $\Delta V$ measured along the line in d at $I = -62$~pA. $\Delta V$ peaks at position of the ligand lobes, indicating the crucial role of the IS. The red line is guide to the eye.
 \label{Fig4}} 
\vspace{1cm}

\end{figure}

Fig.~4a summarizes the sequence of spectra taken along the molecule axis indicated in the central panel of Fig.~3e. By moving towards the edge of the molecule the ionization threshold increases as a direct consequence of the reduced tip-induced gating illustrated in Fig.~4b but always takes place in a voltage range between $-1.5 ... -2$~V. Although the ionization voltage somewhat depends on the exact tip shape (note the slightly different thresholds obtained using different tips but same stabilization parameters in Fig. 2 and 4), our data indicate that this general trend is very robust,  with all molecules displaying the very same behavior (see Supplementary Information) compatible with low voltage electronic devices. 

Overall, the voltage-current relation resembles the transport characteristic of a SET \cite{K1992}, where a gate voltage $V_{\rm g}$ allows to select the discrete number of occupied states determining the current flowing between two terminals, i.e.\ source and drain, once a voltage $V_{\rm sd}$ is applied. In the present case, the bias applied through the tunneling junction correspond to $V_{\rm sd}$, while the tip position mimics a gate potential $V_{\rm g}$ which allows---through capacitive coupling---to regulate the voltage drop at the molecule/TI interface \cite{FKS2012,ZWB2014}. The discrete characteristic is schematically illustrated in Fig.~4c, where differently shaded areas correspond to distinct transconductance states that are separated by Coulomb blockade.  

In order to quantify the extension and the strength of the gating potential created by the additional electron---both important quantities which directly impact macroscopic properties like transport---we further analyzed the steps visible in the tunneling current at the ionization threshold. Fig.~4d shows tunneling spectra measured above a molecule ligand and on the bare TI surface (see red and black points in the topographic image in the left inset of Fig.~4d, respectively). While the $I(V)$ spectra overlap at low negative bias voltage ($V > -1.7$~V), they are shifted by a fixed voltage beyond the ionization threshold, where a sudden step-like increase of the current is visible.

The subtle changes induced in the tunneling process by the molecule's ionization prevent an exact quantification of the its gating potential. Nevertheless,  similar to previous observation on single dopants buried in conventional semiconductors  \cite{TVL2008}, an estimation of the gating potential can be obtained by analyzing the voltage shift $\Delta V$ between the ionized and neutral $I(V)$ curves at a position corresponding to the step-like increase of the tunneling current ($-62$~pA in this case). Results obtained along the line in the left inset of Fig.~4d are reported in Fig.~4e. Values up to $\Delta V = 0.35$~V are observed at a distance of about 1~nm from the center of the molecule. This value is close to the sum of the static band bending induced by the molecule adsorption process, i.e. 70~meV, plus the 300~meV shift required to align the interface state with the Fermi level, and therefore consistent with the flat band condition sketched in Fig.~3d.

Our findings illuminate the interaction of topological states with the localized Coulomb potential of single electron charges. The use of the high degree of charge localization achievable in molecules opens up a new pathway for reliably interfacing TIs with well-defined external perturbations. If correlated with the ionization state of particular chemical species like in the present case, the resulting bistability proves that hybrid molecule/TI interfaces may be suitable as sensors. More generally, the capability to dynamically gate molecules to the single-electron level through electric fields, paves the way towards purely electrically controlled spintronic devices, thereby bringing us a significant step closer towards an application of this fascinating class of materials.

The authors gratefully acknowledge discussions with Michael Flatt\'e.

\vspace{1cm}
{\bf Methods}
Experiments have been performed in an STM operated at $T = 4.8$~K. n-doped Bi$_2$Te$_3$ single-crystals were grown using the Bridgman technique. The crystal structure consists of alternating planes of Bi and Te up to the formation of a quintuple layer with the sequence Te-Bi-Te-Bi-Te. Quintuple layers are weakly coupled by van der Waals forces thus offering a natural cleaving plane. The Bi$_2$Te$_3$ single crystals have been cleaved in UHV at a base pressure of $3 \cdot 10^{-11}$ mbar and immediately inserted into the STM. Because of the above mentioned crystal structure, cleaved crystals are always Te terminated. MnPc molecules (Sigma-Aldrich) were deposited directly onto the cold Bi$_2$Te$_3$ surface by using a home-made Knusden cell and annealed at room temperature. STM measurements have been performed using electrochemically etched tungsten tips. Spectroscopic data have been acquired by lock-in technique ($f=793$ Hz, $V_{rms}$ = 10 mV).

\end{document}